\documentclass[aps,prl,preprint,showpacs,superscriptaddress]{revtex4-1}

\usepackage{bm,graphicx,textcomp,amssymb,amsmath,dcolumn,hyperref,color}

\newcommand{\nn}{\nonumber\\}

\newcommand{\f}[1]{\mbox{\boldmath$#1$}}

\newcommand{\bea}{\begin{eqnarray}}
\newcommand{\ea}{\end{eqnarray}}
\newcommand{\eea}{\end{eqnarray}}
\newcommand{\ord}{\,{\cal O}}

\begin{document}

\title{Giant magneto-photoelectric effect in suspended graphene}

\author{Jens~Sonntag}
\author{Annika~Kurzmann}
\author{Martin~Geller}
\affiliation{Fakult\"at f\"ur Physik and CENIDE, Universit\"at Duisburg-Essen, 
Lotharstra{\ss}e 1, Duisburg 47048, Germany}
\author{Friedemann~Queisser}
\affiliation{Fakult\"at f\"ur Physik, 
Universit\"at Duisburg-Essen, Lotharstra{\ss}e 1, Duisburg 47048, Germany}
\author{Axel~Lorke}
\affiliation{Fakult\"at f\"ur Physik and CENIDE, 
Universit\"at Duisburg-Essen, Lotharstra{\ss}e 1, Duisburg 47048, Germany}
\author{Ralf~Sch\"utzhold}
\affiliation{Fakult\"at f\"ur Physik, 
Universit\"at Duisburg-Essen, Lotharstra{\ss}e 1, Duisburg 47048, Germany}

\date{\today}

\begin{abstract}
We study the optical response of a suspended graphene field-effect
transistor in magnetic fields of up to 9~T (quantum Hall regime).
With an illumination power of only 3~$\mu$W, we measure a photocurrent of up 
to 400~nA, 
corresponding to a photo-responsivity of 0.14~A/W, 
which we believe to be 
the highest value ever measured in single-layer graphene.
We estimate that every absorbed photon creates more than 8 electron-hole 
pairs, which demonstrates highly effective carrier multiplication. 
As suggested by the dependence of the photocurrent on gate voltage and 
magnetic field, 
we propose a ballistic two-stage mechanism where the incident photons 
create primary charge carriers which then excite secondary charge carriers 
in the chiral edge states via Auger-type inelastic Coulomb scattering 
processes at the graphene edge.
\end{abstract}


\maketitle

For many years, the famous paper by W.B.~Shockley and H.-J.~Queisser from 1961 
has been the standard for assessing the maximum efficiency of 
semiconductor solar cells \cite{Shockley61}. 
In brief, the main argument is based on the assumption that photons with 
energies below the band-gap of the semiconductor are not absorbed while the 
excess energy of photons above the band-gap is dissipated as heat and not 
converted into electric energy. 
Recently, however, the Shockley-Queisser limit has been under close scrutiny, 
as some of the limiting factors may be overcome using novel, tailored materials 
and mechanisms, which were not envisioned 50 years ago. 
One such mechanism, which recently received great attention, is carrier 
multiplication \cite{Schaller04,Chan12, Beard07,Wang10,Congreve13,
Tielrooij13,Ploetzing14,Mittendorff15}: 
Charge carriers, which are optically excited with a surplus energy,
can relax by exciting electron-hole pairs, effectively turning a single 
photon into two or more electron-hole pairs that can drive an external circuit.

\begin{figure}
	\includegraphics[scale=1]{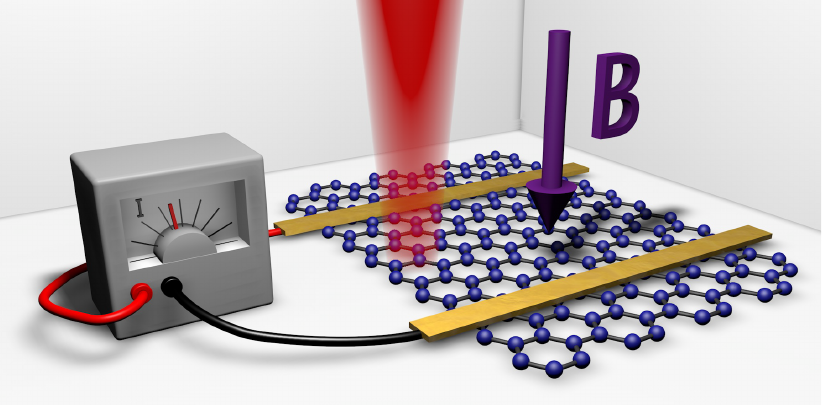}
	\caption{Schematic of the photocurrent measurement.}
	\label{fig1a}
\end{figure}

For optical or infrared absorption, the two-dimensional crystal graphene is a 
promising material, as its pseudo-relativistic energy-momentum relation 
$E(\f{p})\approx v_{\rm F} |\f{p}|$ at low energies 
(where $v_{\rm F}\approx10^6\rm m/s$ is the Fermi velocity) gives rise to a 
broad absorption bandwidth. 
On the other hand, the absence of an energy gap seems to rule out the usual 
mechanism for charge separation in semiconductor solar cells via a built-in 
electrical potential gradient, and graphene $pn$-junctions are challenging to 
fabricate \cite{Gabor11,Williams07}.
Recently, some of us proposed to employ the magneto-photoelectric effect 
along a graphene fold or edge to achieve charge separation \cite{Queisser13}
using an applied magnetic field $B$ instead of an electrical potential 
gradient, see Figure~\ref{fig1a}.

Apart from the high charge carrier mobility and the
broad absorption bandwidth, which remains true for the edge 
states in a magnetic field, graphene has a number of further interesting 
properties for magneto-photocurrent generation.
For example, the cyclotron radius of a pseudo-relativistic electronic 
excitation in graphene 
\bea
\label{cyclotron-radius} 
r=\frac{|\f{p}|}{qB}=\frac{E}{qBv_{\rm F}}
\,,
\ea
with the elementary charge $q$ and an energy $E$ corresponding to, say, 
room temperature  
within a magnetic field $B$ of 4~Tesla (see below) is 6~nm and thus well 
below the mean free path. 
Furthermore, in addition to this classical length scale, the magnetic 
(Landau) length $\ell_B=\sqrt{\hbar/(qB)}\approx13~\rm nm$ is of the same 
order -- which shows that quantum effects have to be taken into account. 
The third advantage of graphene lies in the relatively strong Coulomb 
interaction: 
In analogy to quantum electrodynamics (QED), we can construct an effective 
fine-structure constant in graphene
\bea
\label{fine-structure-constant}
\alpha_{\rm graphene}
=
\frac{q^2}{4\pi\hbar\varepsilon_0v_{\rm F}}
=
\frac{c}{v_{\rm F}}\,\frac{q^2}{4\pi\hbar\varepsilon_0c}
=
\frac{c}{v_{\rm F}}\,\alpha_{\rm QED}
\,,
\ea
and find that this coupling strength $\alpha_{\rm graphene}$ is much larger
than $\alpha_{\rm QED}\approx1/137$ due to $c/v_{\rm F}\approx300$.
Intuitively speaking, the charge carriers are much slower than the speed of 
light $c$ and thus have more time to interact. 
These comparably strong interactions will be important for charge carrier 
multiplication discussed below \cite{foot-absorption}.

\begin{figure}
\includegraphics[scale=1]{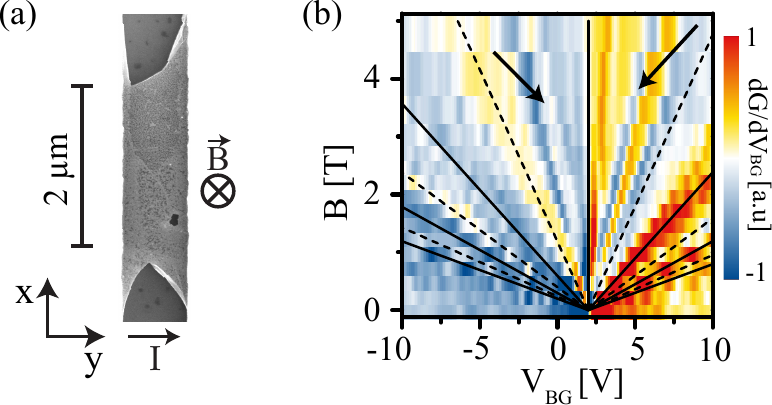}
\caption{(a) Scanning electron micrograph of the suspended graphene structure, 
taken after the measurements. 
The Ti/Au contacts can be seen as bright edges on the left and right hand side. 
(b) Two-terminal transconductance as a function of the magnetic field and the 
gate voltage. 
Dotted (continuous) lines display filled (half filled) Landau levels, 
respectively, calculated from the Landau energy spectrum and the geometric 
capacitance. 
Two arrows show the position of additional plateaus, indicating the splitting 
of the zeroth level.}
\label{fig1b}
\end{figure}

Motivated by the predictions in \cite{Queisser13},
here we experimentally investigate the photocurrent generation in suspended 
graphene in a quantizing magnetic field. 
We start from commercially available CVD graphene, transferred to a 285 nm 
SiO$_{2}$-on-Si substrate. 
The silicon substrate is highly doped and is used as a back gate electrode. 
Using photolithography and an oxygen plasma, the graphene is patterned into 
bars of $\approx2.5~\mu$m width.
Afterwards, Ti/Au (5/100~nm) electrodes are defined by electron beam 
lithography. 
The contacts cover the whole width of the graphene bars and are separated 
by 660~nm. 
In a subsequent etching step using hydrofluoric acid, 
$\approx160$~nm of SiO$_{2}$ 
are removed below the graphene, which creates a suspended field-effect 
transistor structure~\cite{Bolotin08,Sommer15}.  
A scanning electron micrograph of the suspended graphene 
(taken after the photocurrent measurements) is shown in Fig.~\ref{fig1b}~(a).

Our measurement setup consists of a confocal microscope inside a liquid helium 
cryostat, which allows us to measure at a temperature of 4.2~K and in 
magnetic fields of up to 9~T. 
After cooling down, the graphene is cleaned \textit{in-situ} by a current 
annealing step \cite{Moser07,Bolotin08}. 
To check the quality of the graphene, we perform transconductance measurements
d$G$/d$V_{BG}$ in different magnetic fields. 
Here $G$ is the conductance of the graphene channel and $V_{BG}$ is the bias 
applied to the back gate to control the charge carrier density in the graphene.
The resulting Shubnikov-de~Haas oscillations
are shown in Fig.~\ref{fig1b}~(b). 
The Dirac point  
lies at $V_{BG}=2~\mathrm{V}$, 
so that the Landau levels fan out from this point. 
The expected fan diagram [dotted and solid lines in Fig.~\ref{fig1b}~(b)] is 
calculated from the geometric capacitance and the theoretical energy spectrum 
of single layer graphene $E_j=\pm\sqrt{2q\hbar v_{\rm F}^2 B\left|j\right|}$, 
where $j$ is the Landau level index. 
The good agreement with the experimental data confirms the geometric 
capacitance and shows that the graphene is suspended in a single layer. 
The Shubnikov-de~Haas oscillations can be observed at magnetic fields 
as low as 0.5~T, indicating a mobility of 
$\mu>20~000~\mathrm{cm^2/(Vs)}$ \cite{Bolotin08}. 
At magnetic fields above 1 T, additional plateaus appear, showing at least partial 
splitting of the zeroth Landau level. 
This effect is often observed in high-quality samples and is frequently 
attributed to electron-electron interactions \cite{Du09,Bolotin09,Young12}. 

\begin{figure}
\includegraphics[scale=1]{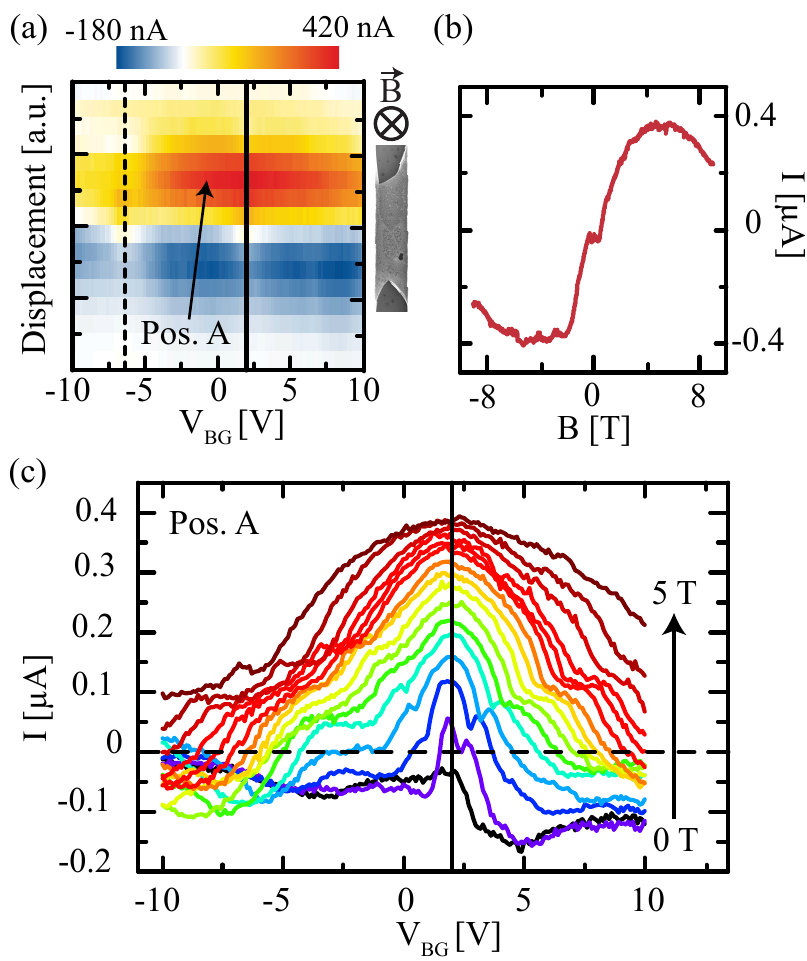}
\caption{(a) Photocurrent at $B=5$~T as a function of the laser spot position 
and the gate voltage. 
The color scale is linear, with white corresponding to 0~nA. 
The solid vertical line shows the Dirac point,  
the dashed line indicates an empty zeroth Landau level. 
The picture of the graphene on the right hand side roughly indicates the position 
of the laser spot. 
(b) photocurrent of position A at the Dirac point  
as a function of the magnetic field. 
(c) Gate voltage dependence of the photocurrent for different magnetic fields.}
\label{fig2}
\end{figure}

Next we investigate the photocurrent generation in suspended graphene. 
For this purpose, we use the confocal microscope and a near-infrared laser 
with a wavelength of 972~nm and a spot diameter of roughly 1.5~$\mu$m. 
The illumination power is 3 $\mu$W and we use a low-impedance ($<60~\Omega$) 
current amplifier to directly measure the photocurrent. 
The laser spot is positioned in between the two gold contacts. 
Since the distance between the contacts is less than the spot diameter, 
only the spatial dependence perpendicular to the current flow can be resolved. 
Figure~\ref{fig2}~(a) shows the generated photocurrent at $B=5$~T, 
well within the quantum Hall regime. 
The photocurrent is found to be symmetric around the Dirac point, 
but there is a clear change in polarity, depending on which edge of
the graphene is illuminated. 
This is in agreement with the magnetic-field induced chirality of the 
charge-carrier motion (see below).  

As Fig.~\ref{fig2}~(a) shows, the measured photocurrent reaches surprisingly 
high values of over 400~nA at the Dirac point when illuminating the upper 
edge  
[position A in Fig.~\ref{fig2}~(a)]. 
Considering the illumination power of 3~$\mu$W, this corresponds to a 
photo-responsivity of 0.14~A/W, a value more than an order of magnitude 
higher than reported previously for single layer graphene devices 
\cite{Nazin10, Freitag13} and comparable to the responsivity of 
commercially available photodiodes. 
Assuming that the entire laser spot illuminates graphene and taking into 
account the absorption of graphene of $\approx 2.3\% $ \cite{Nair08}, 
this high current of 400~nA implies that for every absorbed photon 8 
electron-hole pairs are created. 
We believe that this is 
the highest value of charge carrier multiplication observed in graphene 
as well as in any optically excited semiconductor system so far 
\cite{Schaller04,Chan12, Beard07, Wang10,Congreve13,Tielrooij13,
Ploetzing14,Mittendorff15}.

The gate-voltage dependence of the magneto-photocurrent at different magnetic 
fields is shown in Fig.~\ref{fig2}~(c). 
The photocurrent at zero magnetic field (black line) features a step at the 
Dirac point,  
often observed at $pn$ or graphene-metal junctions, 
and commonly 
attributed to thermoelectric effects \cite{Gabor11}. 
For an applied magnetic field, the step at the Dirac point 
turns into a peak, whose height and width first increases with 
increasing 
magnetic fields but later (for $B>5~$T) decreases again, 
see Fig.~\ref{fig2}~(b) \cite{foot1}. 
This is in strong contrast to the expected oscillatory behavior of 
usual (diffusive) thermoelectric effects in the quantum Hall 
regime \cite{Zuev09,Checkelsky09,Jonson84,Girvin82}, see also \cite{foot-nazin}.
Furthermore, we find a distinct polarization dependence 
(not shown here \cite{Sonntag14}), which is also not expected for a 
thermoelectric response \cite{foot2}.
\begin{figure}
\includegraphics[scale=1]{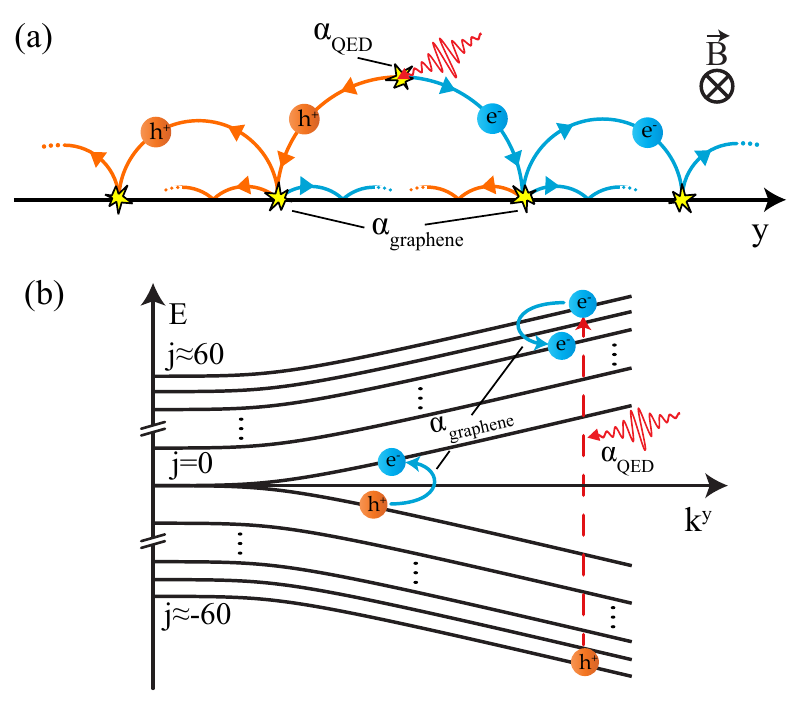}
\caption{
Sketch of the proposed mechanism in real space (a) and in an energy-momentum 
diagram (b). 
In a semi-classical picture (a), after optical excitation 
(with $\alpha_{\rm QED}$), the generated primary charge carriers move along 
curved trajectories until they collide with the graphene edge 
(horizontal black line), at which point they can excite 
secondary electron-hole pairs through inelastic Coulomb scattering
(governed by $\alpha_{\rm graphene}$). 
This Auger-type process is depicted in (b) within the energy dispersion 
diagram of the edge channels.}
\label{fig3}
\end{figure}

Based on these observations and our previous work \cite{Queisser13}, 
we propose the following two-stage ballistic mechanism (sketched in Fig.~\ref{fig3}) 
based on the creation of primary and secondary charge carriers and their 
transport far from equilibrium:
With the universal absorption probability $\pi\alpha_{\rm QED}\approx2.3~\%$,
the incident photon of energy $\approx1.2~$eV creates a pair of an electron 
and a hole with equal energies of $E\approx0.6~\rm eV$ and opposite momenta due to 
energy-momentum conservation.
However, the momentum of the electron (or hole) can point in both directions 
with equal probability and thus no net current is generated at this stage.
Since the wavelength of the electron or hole excitations 
$\lambda\approx7~\rm nm$ is much smaller than the classical cyclotron 
radius~(\ref{cyclotron-radius}) of approximately 150~nm at 4~T,
we can treat the propagation semi-classically.
Thus, the electron and hole excitations describe circular trajectories with 
the cyclotron radius~(\ref{cyclotron-radius}) until they reach the metallic
contacts or they are scattered by defects or the graphene edge. 

A net current is induced when at least one of the carriers is reflected at the 
graphene edge, where the originally random direction of charge separation 
is transformed into a determined directionality/chirality as in 
Figure~\ref{fig3}, where holes move to the left and electrons to the right.
The current runs into opposite directions at the upper and lower edge, 
which explains the position dependence in Fig.~\ref{fig2}~(a). 
This simple picture also accounts for the observed dependence on the magnetic 
field: 
If the magnetic field is too small, the radius~(\ref{cyclotron-radius}) is 
much larger than the distance between the metallic contacts and thus the 
trajectories are not bent enough to 
control (rectify) the direction of
charge separation efficiently.
For intermediate field strengths of around 4~T, the circular diameter of 
300~nm fits well into the graphene sample and thus 
 (directed) charge separation is most efficient.
If the magnetic field becomes too large, however, this diameter shrinks and 
thus the incident photon must be absorbed very near the edge when the circle 
is supposed to intercept the edge -- i.e., the effective absorption 
area shrinks. 
However, as explained above, these primary electron-hole pairs 
(directly created by the incident laser photons) 
cannot account for the observed current.
The observed carrier multiplication can be explained by inelastic scattering 
at the graphene edge.
Neglecting dielectric and screening effects in our 
order-of-magnitude estimate, we consider the Coulomb interaction Hamiltonian 
\bea
\label{Coulomb}
\hat H_{\rm Coulomb}=\frac{q^2}{2}\int d^2 r \int d^2 r'\,
\frac{\hat\varrho(\f{r})\hat\varrho(\f{r'})}{4\pi\varepsilon_0|\f{r}-\f{r'}|}  
\,,
\ea
with the charge density operator 
$\hat\varrho(\f{r})=\hat{\f{\psi}}^\dagger(\f{r})\cdot\hat{\f{\psi}}(\f{r})$, 
where $\hat{\f{\psi}}(\f{r})$ is the two-component (spinor) field operator. 
In principle, this non-linear interaction Hamiltonian could also induce 
carrier multiplication in translationally invariant (bulk) graphene, but 
this process is strongly suppressed due to energy-momentum conservation
\cite{foot-phase-space}. 
This supression can be diminished by the coupling to phonons, defects, or a 
magnetic field, etc., and carrier multiplication has been observed in such 
scenarios \cite{Tielrooij13,Ploetzing14, Mittendorff15, Wendler14}. 

Here, we consider the inelastic reflection at the graphene edge 
(see Fig.~\ref{fig3}) where the perpendicular momentum is not conserved 
and thus we expect strong carrier multiplication effects. 
Since the primary electron or hole excitation has a comparably large 
$k_{\rm in}\approx2\pi/(7~\rm nm)$, we approximate it by a plane wave 
$\f{\psi}_{\rm in}\sim\exp\{i\f{k}_{\rm in}\cdot\f{r}\}$ 
and similarly for the outgoing excitation $\f{\psi}_{\rm out}$ 
after inelastic scattering at the edge. 
In order to deal with well-defined modes for the secondary excitations, 
we consider the fold modes discussed in \cite{Queisser13}, but other models 
yield very similar results. 
Furthermore, a closer inspection of Fig.~\ref{fig1b}~(a) suggests that the 
graphene edge is indeed folded (at least over a wide range) such that the 
fold modes should be a good description. 
Assuming that the edge runs along the $y$-axis, in the lowest band, these modes behave as 
$\f{\psi}_\pm\sim\exp\{ik_\pm^yy-\Omega_\pm x^2/2\}$ for large $k$, 
where $\f{\psi}_+$ denotes the electron and $\f{\psi}_-$ the hole excitation. 
These modes are localized (in $x$-direction) at the fold and propagate 
along it (in $y$-direction) with nearly $\pm v_{\rm F}$.
Their transverse extent is determined by 
$\Omega_\pm^{-1}=\ell_B(R/k_\pm^y)^{1/2}$ where $R$ is the curvature radius of the 
graphene fold (roughly between 10 and 20~nm). 

Via standard perturbation theory with respect to the interaction 
Hamiltonian~(\ref{Coulomb}), we calculate the amplitude (matrix element) for 
Auger-type inelastic scattering of a primary excitation from $\f{k}_{\rm in}$ 
to $\f{k}_{\rm out}$ while creating a secondary electron-hole pair with 
$k_\pm^y$. 
The conservation of energy and momentum in $y$-direction determines the 
wave-numbers $k_\pm^y$ uniquely and the remaining matrix element reads 
(for large $k$)
\bea
{\cal M}
&=&
\frac{q^2}{4\pi\hbar\varepsilon_0v_{\rm F}}\,
\frac{(\Omega_+\Omega_-)^{1/4}}{(\Omega_++\Omega_-)^{1/2}}\,
\exp\left\{-\frac{(k_{\rm in}^x-k_{\rm out}^x)^2}{2(\Omega_++\Omega_-)}\right\}
\nn
&&
\times
\frac{\left[k_{\rm in}^x+i(k_{\rm in}+k_{\rm in}^y)\right]
\left[k_{\rm out}^x+i(k_{\rm out}-k_{\rm out}^y)\right]}
{4k_{\rm in}k_{\rm out}
\sqrt{\Delta k_y^2+(\Delta k_x+\kappa)^2}
}
\,,
\ea
where $\Delta\f{k}=(k_+^y\f{e}_y+k_-^y\f{e}_y-\f{k}_{\rm in}-\f{k}_{\rm out})/2$ is the 
momentum transfer and 
$\kappa=\frac12(k_{\rm in}^x-k_{\rm out}^x)(\Omega_+-\Omega_-)/(\Omega_++\Omega_-)$
quantifies the particle-hole asymmetry.
In the perpendicular case $k_{\rm in}^y=k_{\rm out}^y=0$, 
we get $k_+^y=k_-^y\approx(k_{\rm in}-k_{\rm out})/2$ and thus 
$\Omega_+=\Omega_-$ such that this asymmetry disappears $\kappa=0$ and the 
above formula simplifies strongly.
The total probability for this Auger-type carrier multiplication process can 
then be estimated by
\bea
{\cal P}
=
\int d^2k_{\rm out}\,\left|{\cal M}\right|^2
=
\ord\left(\left[
\frac{q^2}{4\pi\hbar\varepsilon_0v_{\rm F}}
\right]^2\right)
\,,
\ea
which yields the squared coupling constant $\alpha_{\rm graphene}^2$ in 
Eq.~(\ref{fine-structure-constant}). 

This general scaling ${\cal P}\sim\alpha_{\rm graphene}^2$ also persists in other 
scenarios or models -- as long as the orders of magnitude of the involved 
length scales are not too far away from the magnetic length $\ell_B$ 
(as it is the case here).
We may understand the rough order of magnitude of the amplitude as the 
product of a characteristic Coulomb interaction energy 
$q^2/(4\pi\varepsilon_0\ell_B)$ and a typical interaction time 
$\ell_B/v_{\rm F}$ (divided by $\hbar$). 
As anticipated in the Introduction, 
the comparably slow Fermi velocity $v_{\rm F}\approx c/300$ implies that the 
probability ${\cal P}$ is enhanced by a huge factor 
$(c/v_{\rm F})^2=\ord(10^5)$ in comparison to ordinary QED.
Strictly speaking, this enhancement already indicates that first-order 
perturbation theory is not really applicable anymore -- but in any case 
we find that the probability ${\cal P}$ is greatly enhanced 
(consistent with our experimental observations).

The presented theoretical model -- ranging from the creation of the primary 
electron-hole excitations by the incident photons and their subsequent 
propagation on circular orbits up to the generation of secondary electron-hole 
excitations via inelastic (Auger-type) scattering at the graphene edge/fold -- 
can explain the observations, including their dependence on laser spot position and magnetic field, 
quite well.  
The gate-voltage dependence can also be understood within this picture:
Deviations from the Dirac point reduce the available phase space for the 
generation of secondary electron-hole excitations since the required energy 
increases. 
Note that only Auger-type electron-hole excitations from a downward to an 
upward sloping dispersion curve contribute to the net current because the 
current is determined by the group velocity $dE/dk^y$ in 
Fig.~\ref{fig3}~(b), see \cite{Queisser13}.

In summary, we have observed a surprisingly high magneto-photocurrent in 
suspended graphene. 
The measured current indicates that for every absorbed photon, 
more than eight electron-hole pairs are generated. 
The experimental results can be explained by a theoretical model based 
on charge carrier multiplication via inelastic scattering of primary 
photo-generated excitations at the graphene edge.
The effectiveness of this process can mainly be attributed to three factors:
a) the comparably strong Coulomb interaction quantified by 
$\alpha_{\rm graphene}\gg\alpha_{\rm QED}$,
b) the enlarged phase space due to the lifting of the transversal 
momentum conservation at the edge,
c) the robust chiral edge states in a magnetic field,  
which ensure that basically all secondary carriers generated at the 
edge will contribute to the photocurrent. 
Our findings show that graphene with its high carrier mobility and 
broad absorption bandwidth is a very promising material for 
photo-electric applications. 

\acknowledgments

R.S.\ and F.Q.\ were supported by DFG (SFB-TR12).


\begin{thebibliography}{99}
\bibitem{foot-absorption}
	Note that $\alpha_{\rm graphene}$ corresponds to the Coulomb interaction 
	between the electronic excitations, their coupling to photons 
	(e.g., the universal light absorption $\approx2.3~\%$ of graphene) 
	is still governed by $\alpha_{\rm QED}$ 
	(or, more precisely $\pi\alpha_{\rm QED}$ \cite{Nair08}).
\bibitem{foot1}
	The photocurrent of the other edge [blue area in Fig.~\ref{fig2}~(a)] 
	shows stronger oscillations and thus more similarity to the well-known 
	magneto-thermoelectric effects. 
	This may result from the particular geometrical shapes or different disorder 
	and defect densities of the two edges.
\bibitem{foot2}
	Detailed calculations of the optical field distribution at the location of the graphene stripe have shown that a possible polarizing effect of the two parallel, metallic contact electrodes can be ruled out as a reason for the polarization dependence \cite{Sonntag14}.
\bibitem{foot-phase-space}
	Considering the case where an electron is scattered inelastically 
	from $\f{k}_{\rm in}$ to $\f{k}_{\rm out}$ while creating a secondary 
	electron-hole pair with $\f{k}_+$ and $\f{k}_-$, energy-momentum conservation 
	together with the linear dispersion relation 
	$E(\f{k})\approx\hbar v_{\rm F}|\f{k}|$ implies that all these wavenumbers 
	must be parallel $\f{k}_{\rm in}\|\f{k}_{\rm out}\|\f{k}_+\|\f{k}_-$ 
	such that the phase-space volume is basically zero.
\end{thebibliography}
\end{document}